\documentstyle[pra,aps,psfig,epsfig]{revtex}
\begin{document}
\draft
\title{
\begin{flushright}
{\bf Preprint SSU-HEP-02/04\\
Samara State University}
\end{flushright}
\vspace{5mm}
\begin{center}
PSEUDOSCALAR POLE TERMS CONTRIBUTIONS\\
TO HADRONIC LIGHT-BY-LIGHT CORRECTIONS\\
TO THE MUONIUM HYPERFINE SPLITTING
\end{center}
}
\author{R.N.Faustov\footnote{E-mail:
faustov@theory.sinp.msu.ru}}
\address{117333, Moscow, Vavilov, 40, Scientific Council "Cybernetics" RAS }
\author{A.P.Martynenko\footnote{E-mail:
mart@info.ssu.samara.ru}}
\address{443011, Samara, Pavlov, 1, Department of Theoretical Physics,
Samara State University}

\date{\today}

\maketitle

\begin{abstract}
The contribution of pseudoscalar pole diagrams to the
hadronic light-by-light corrections of order ${\rm\alpha^3 E_F}$
to the ground state hyperfine splitting in hydrogenic atom is calculated
in the pseudoscalar pole terms approximation. The vector dominance
model for the form factor of the transition of a pseudoscalar meson into two
photons is used. With the account of new experimental data on the cross
sections ${\rm\sigma (e^+e^-\rightarrow \rho, \omega \rightarrow
\pi^0(\eta)\gamma)}$ in the SND and CMD-2 experiments some other hadronic
corrections to the muonium hyperfine splitting are calculated.
\end{abstract}

\pacs{36.10.Dr, 12.20.Ds, 14.40.Aq, 12.40.Vv}

\immediate\write16{<<WARNING: LINEDRAW macros work with emTeX-dvivers
                    and other drivers supporting emTeX \special's
                    (dviscr, dvihplj, dvidot, dvips, dviwin, etc.) >>}
\newdimen\Lengthunit       \Lengthunit  = 1.5cm
\newcount\Nhalfperiods     \Nhalfperiods= 9
\newcount\magnitude        \magnitude = 1000

\catcode`\*=11
\newdimen\L*   \newdimen\d*   \newdimen\d**
\newdimen\dm*  \newdimen\dd*  \newdimen\dt*
\newdimen\a*   \newdimen\b*   \newdimen\c*
\newdimen\a**  \newdimen\b**
\newdimen\xL*  \newdimen\yL*
\newdimen\rx*  \newdimen\ry*
\newdimen\tmp* \newdimen\linwid*

\newcount\k*   \newcount\l*   \newcount\m*
\newcount\k**  \newcount\l**  \newcount\m**
\newcount\n*   \newcount\dn*  \newcount\r*
\newcount\N*   \newcount\*one \newcount\*two  \*one=1 \*two=2
\newcount\*ths \*ths=1000
\newcount\angle*  \newcount\q*  \newcount\q**
\newcount\angle** \angle**=0
\newcount\sc*     \sc*=0

\newtoks\cos*  \cos*={1}
\newtoks\sin*  \sin*={0}

\catcode`\[=13

\def\rotate(#1){\advance\angle**#1\angle*=\angle**
\q**=\angle*\ifnum\q**<0\q**=-\q**\fi
\ifnum\q**>360\q*=\angle*\divide\q*360\multiply\q*360\advance\angle*-\q*\fi
\ifnum\angle*<0\advance\angle*360\fi\q**=\angle*\divide\q**90\q**=\q**
\def\sgcos*{+}\def\sgsin*{+}\relax
\ifcase\q**\or
 \def\sgcos*{-}\def\sgsin*{+}\or
 \def\sgcos*{-}\def\sgsin*{-}\or
 \def\sgcos*{+}\def\sgsin*{-}\else\fi
\q*=\q**
\multiply\q*90\advance\angle*-\q*
\ifnum\angle*>45\sc*=1\angle*=-\angle*\advance\angle*90\else\sc*=0\fi
\def[##1,##2]{\ifnum\sc*=0\relax
\edef\cs*{\sgcos*.##1}\edef\sn*{\sgsin*.##2}\ifcase\q**\or
 \edef\cs*{\sgcos*.##2}\edef\sn*{\sgsin*.##1}\or
 \edef\cs*{\sgcos*.##1}\edef\sn*{\sgsin*.##2}\or
 \edef\cs*{\sgcos*.##2}\edef\sn*{\sgsin*.##1}\else\fi\else
\edef\cs*{\sgcos*.##2}\edef\sn*{\sgsin*.##1}\ifcase\q**\or
 \edef\cs*{\sgcos*.##1}\edef\sn*{\sgsin*.##2}\or
 \edef\cs*{\sgcos*.##2}\edef\sn*{\sgsin*.##1}\or
 \edef\cs*{\sgcos*.##1}\edef\sn*{\sgsin*.##2}\else\fi\fi
\cos*={\cs*}\sin*={\sn*}\global\edef\gcos*{\cs*}\global\edef\gsin*{\sn*}}\relax
\ifcase\angle*[9999,0]\or
[999,017]\or[999,034]\or[998,052]\or[997,069]\or[996,087]\or
[994,104]\or[992,121]\or[990,139]\or[987,156]\or[984,173]\or
[981,190]\or[978,207]\or[974,224]\or[970,241]\or[965,258]\or
[961,275]\or[956,292]\or[951,309]\or[945,325]\or[939,342]\or
[933,358]\or[927,374]\or[920,390]\or[913,406]\or[906,422]\or
[898,438]\or[891,453]\or[882,469]\or[874,484]\or[866,499]\or
[857,515]\or[848,529]\or[838,544]\or[829,559]\or[819,573]\or
[809,587]\or[798,601]\or[788,615]\or[777,629]\or[766,642]\or
[754,656]\or[743,669]\or[731,681]\or[719,694]\or[707,707]\or
\else[9999,0]\fi}

\catcode`\[=12

\def\GRAPH(hsize=#1)#2{\hbox to #1\Lengthunit{#2\hss}}

\def\Linewidth#1{\global\linwid*=#1\relax
\global\divide\linwid*10\global\multiply\linwid*\mag
\global\divide\linwid*100\special{em:linewidth \the\linwid*}}

\Linewidth{.4pt}
\def\sm*{\special{em:moveto}}
\def\sl*{\special{em:lineto}}
\let\moveto=\sm*
\let\lineto=\sl*
\newbox\spm*   \newbox\spl*
\setbox\spm*\hbox{\sm*}
\setbox\spl*\hbox{\sl*}

\def\mov#1(#2,#3)#4{\rlap{\L*=#1\Lengthunit
\xL*=#2\L* \yL*=#3\L*
\xL*=\xscale\xL* \yL*=\yscale\yL*
\rx* \the\cos*\xL* \tmp* \the\sin*\yL* \advance\rx*-\tmp*
\ry* \the\cos*\yL* \tmp* \the\sin*\xL* \advance\ry*\tmp*
\kern\rx*\raise\ry*\hbox{#4}}}

\def\rmov*(#1,#2)#3{\rlap{\xL*=#1\yL*=#2\relax
\rx* \the\cos*\xL* \tmp* \the\sin*\yL* \advance\rx*-\tmp*
\ry* \the\cos*\yL* \tmp* \the\sin*\xL* \advance\ry*\tmp*
\kern\rx*\raise\ry*\hbox{#3}}}

\def\lin#1(#2,#3){\rlap{\sm*\mov#1(#2,#3){\sl*}}}

\def\arr*(#1,#2,#3){\rmov*(#1\dd*,#1\dt*){\sm*
\rmov*(#2\dd*,#2\dt*){\rmov*(#3\dt*,-#3\dd*){\sl*}}\sm*
\rmov*(#2\dd*,#2\dt*){\rmov*(-#3\dt*,#3\dd*){\sl*}}}}

\def\arrow#1(#2,#3){\rlap{\lin#1(#2,#3)\mov#1(#2,#3){\relax
\d**=-.012\Lengthunit\dd*=#2\d**\dt*=#3\d**
\arr*(1,10,4)\arr*(3,8,4)\arr*(4.8,4.2,3)}}}

\def\arrlin#1(#2,#3){\rlap{\L*=#1\Lengthunit\L*=.5\L*
\lin#1(#2,#3)\rmov*(#2\L*,#3\L*){\arrow.1(#2,#3)}}}

\def\dasharrow#1(#2,#3){\rlap{{\Lengthunit=0.9\Lengthunit
\dashlin#1(#2,#3)\mov#1(#2,#3){\sm*}}\mov#1(#2,#3){\sl*
\d**=-.012\Lengthunit\dd*=#2\d**\dt*=#3\d**
\arr*(1,10,4)\arr*(3,8,4)\arr*(4.8,4.2,3)}}}

\def\clap#1{\hbox to 0pt{\hss #1\hss}}

\def\ind(#1,#2)#3{\rlap{\L*=.1\Lengthunit
\xL*=#1\L* \yL*=#2\L*
\rx* \the\cos*\xL* \tmp* \the\sin*\yL* \advance\rx*-\tmp*
\ry* \the\cos*\yL* \tmp* \the\sin*\xL* \advance\ry*\tmp*
\kern\rx*\raise\ry*\hbox{\lower2pt\clap{$#3$}}}}

\def\sh*(#1,#2)#3{\rlap{\dm*=\the\n*\d**
\xL*=\xscale\dm* \yL*=\yscale\dm* \xL*=#1\xL* \yL*=#2\yL*
\rx* \the\cos*\xL* \tmp* \the\sin*\yL* \advance\rx*-\tmp*
\ry* \the\cos*\yL* \tmp* \the\sin*\xL* \advance\ry*\tmp*
\kern\rx*\raise\ry*\hbox{#3}}}

\def\calcnum*#1(#2,#3){\a*=1000sp\b*=1000sp\a*=#2\a*\b*=#3\b*
\ifdim\a*<0pt\a*-\a*\fi\ifdim\b*<0pt\b*-\b*\fi
\ifdim\a*>\b*\c*=.96\a*\advance\c*.4\b*
\else\c*=.96\b*\advance\c*.4\a*\fi
\k*\a*\multiply\k*\k*\l*\b*\multiply\l*\l*
\m*\k*\advance\m*\l*\n*\c*\r*\n*\multiply\n*\n*
\dn*\m*\advance\dn*-\n*\divide\dn*2\divide\dn*\r*
\advance\r*\dn*
\c*=\the\Nhalfperiods5sp\c*=#1\c*\ifdim\c*<0pt\c*-\c*\fi
\multiply\c*\r*\N*\c*\divide\N*10000}

\def\dashlin#1(#2,#3){\rlap{\calcnum*#1(#2,#3)\relax
\d**=#1\Lengthunit\ifdim\d**<0pt\d**-\d**\fi
\divide\N*2\multiply\N*2\advance\N*\*one
\divide\d**\N*\sm*\n*\*one\sh*(#2,#3){\sl*}\loop
\advance\n*\*one\sh*(#2,#3){\sm*}\advance\n*\*one
\sh*(#2,#3){\sl*}\ifnum\n*<\N*\repeat}}

\def\dashdotlin#1(#2,#3){\rlap{\calcnum*#1(#2,#3)\relax
\d**=#1\Lengthunit\ifdim\d**<0pt\d**-\d**\fi
\divide\N*2\multiply\N*2\advance\N*1\multiply\N*2\relax
\divide\d**\N*\sm*\n*\*two\sh*(#2,#3){\sl*}\loop
\advance\n*\*one\sh*(#2,#3){\kern-1.48pt\lower.5pt\hbox{\rm.}}\relax
\advance\n*\*one\sh*(#2,#3){\sm*}\advance\n*\*two
\sh*(#2,#3){\sl*}\ifnum\n*<\N*\repeat}}

\def\shl*(#1,#2)#3{\kern#1#3\lower#2#3\hbox{\unhcopy\spl*}}

\def\trianglin#1(#2,#3){\rlap{\toks0={#2}\toks1={#3}\calcnum*#1(#2,#3)\relax
\dd*=.57\Lengthunit\dd*=#1\dd*\divide\dd*\N*
\divide\dd*\*ths \multiply\dd*\magnitude
\d**=#1\Lengthunit\ifdim\d**<0pt\d**-\d**\fi
\multiply\N*2\divide\d**\N*\sm*\n*\*one\loop
\shl**{\dd*}\dd*-\dd*\advance\n*2\relax
\ifnum\n*<\N*\repeat\n*\N*\shl**{0pt}}}

\def\wavelin#1(#2,#3){\rlap{\toks0={#2}\toks1={#3}\calcnum*#1(#2,#3)\relax
\dd*=.23\Lengthunit\dd*=#1\dd*\divide\dd*\N*
\divide\dd*\*ths \multiply\dd*\magnitude
\d**=#1\Lengthunit\ifdim\d**<0pt\d**-\d**\fi
\multiply\N*4\divide\d**\N*\sm*\n*\*one\loop
\shl**{\dd*}\dt*=1.3\dd*\advance\n*\*one
\shl**{\dt*}\advance\n*\*one
\shl**{\dd*}\advance\n*\*two
\dd*-\dd*\ifnum\n*<\N*\repeat\n*\N*\shl**{0pt}}}

\def\w*lin(#1,#2){\rlap{\toks0={#1}\toks1={#2}\d**=\Lengthunit\dd*=-.12\d**
\divide\dd*\*ths \multiply\dd*\magnitude
\N*8\divide\d**\N*\sm*\n*\*one\loop
\shl**{\dd*}\dt*=1.3\dd*\advance\n*\*one
\shl**{\dt*}\advance\n*\*one
\shl**{\dd*}\advance\n*\*one
\shl**{0pt}\dd*-\dd*\advance\n*1\ifnum\n*<\N*\repeat}}

\def\l*arc(#1,#2)[#3][#4]{\rlap{\toks0={#1}\toks1={#2}\d**=\Lengthunit
\dd*=#3.037\d**\dd*=#4\dd*\dt*=#3.049\d**\dt*=#4\dt*\ifdim\d**>10mm\relax
\d**=.25\d**\n*\*one\shl**{-\dd*}\n*\*two\shl**{-\dt*}\n*3\relax
\shl**{-\dd*}\n*4\relax\shl**{0pt}\else
\ifdim\d**>5mm\d**=.5\d**\n*\*one\shl**{-\dt*}\n*\*two
\shl**{0pt}\else\n*\*one\shl**{0pt}\fi\fi}}

\def\d*arc(#1,#2)[#3][#4]{\rlap{\toks0={#1}\toks1={#2}\d**=\Lengthunit
\dd*=#3.037\d**\dd*=#4\dd*\d**=.25\d**\sm*\n*\*one\shl**{-\dd*}\relax
\n*3\relax\sh*(#1,#2){\xL*=\xscale\dd*\yL*=\yscale\dd*
\kern#2\xL*\lower#1\yL*\hbox{\sm*}}\n*4\relax\shl**{0pt}}}

\def\shl**#1{\c*=\the\n*\d**\d*=#1\relax
\a*=\the\toks0\c*\b*=\the\toks1\d*\advance\a*-\b*
\b*=\the\toks1\c*\d*=\the\toks0\d*\advance\b*\d*
\a*=\xscale\a*\b*=\yscale\b*
\rx* \the\cos*\a* \tmp* \the\sin*\b* \advance\rx*-\tmp*
\ry* \the\cos*\b* \tmp* \the\sin*\a* \advance\ry*\tmp*
\raise\ry*\rlap{\kern\rx*\unhcopy\spl*}}

\def\wlin*#1(#2,#3)[#4]{\rlap{\toks0={#2}\toks1={#3}\relax
\c*=#1\l*\c*\c*=.01\Lengthunit\m*\c*\divide\l*\m*
\c*=\the\Nhalfperiods5sp\multiply\c*\l*\N*\c*\divide\N*\*ths
\divide\N*2\multiply\N*2\advance\N*\*one
\dd*=.002\Lengthunit\dd*=#4\dd*\multiply\dd*\l*\divide\dd*\N*
\divide\dd*\*ths \multiply\dd*\magnitude
\d**=#1\multiply\N*4\divide\d**\N*\sm*\n*\*one\loop
\shl**{\dd*}\dt*=1.3\dd*\advance\n*\*one
\shl**{\dt*}\advance\n*\*one
\shl**{\dd*}\advance\n*\*two
\dd*-\dd*\ifnum\n*<\N*\repeat\n*\N*\shl**{0pt}}}

\def\wavebox#1{\setbox0\hbox{#1}\relax
\a*=\wd0\advance\a*14pt\b*=\ht0\advance\b*\dp0\advance\b*14pt\relax
\hbox{\kern9pt\relax
\rmov*(0pt,\ht0){\rmov*(-7pt,7pt){\wlin*\a*(1,0)[+]\wlin*\b*(0,-1)[-]}}\relax
\rmov*(\wd0,-\dp0){\rmov*(7pt,-7pt){\wlin*\a*(-1,0)[+]\wlin*\b*(0,1)[-]}}\relax
\box0\kern9pt}}

\def\rectangle#1(#2,#3){\relax
\lin#1(#2,0)\lin#1(0,#3)\mov#1(0,#3){\lin#1(#2,0)}\mov#1(#2,0){\lin#1(0,#3)}}

\def\dashrectangle#1(#2,#3){\dashlin#1(#2,0)\dashlin#1(0,#3)\relax
\mov#1(0,#3){\dashlin#1(#2,0)}\mov#1(#2,0){\dashlin#1(0,#3)}}

\def\waverectangle#1(#2,#3){\L*=#1\Lengthunit\a*=#2\L*\b*=#3\L*
\ifdim\a*<0pt\a*-\a*\def\x*{-1}\else\def\x*{1}\fi
\ifdim\b*<0pt\b*-\b*\def\y*{-1}\else\def\y*{1}\fi
\wlin*\a*(\x*,0)[-]\wlin*\b*(0,\y*)[+]\relax
\mov#1(0,#3){\wlin*\a*(\x*,0)[+]}\mov#1(#2,0){\wlin*\b*(0,\y*)[-]}}

\def\calcparab*{\ifnum\n*>\m*\k*\N*\advance\k*-\n*\else\k*\n*\fi
\a*=\the\k* sp\a*=10\a*\b*\dm*\advance\b*-\a*\k*\b*
\a*=\the\*ths\b*\divide\a*\l*\multiply\a*\k*
\divide\a*\l*\k*\*ths\r*\a*\advance\k*-\r*\dt*=\the\k*\L*}

\def\arcto#1(#2,#3)[#4]{\rlap{\toks0={#2}\toks1={#3}\calcnum*#1(#2,#3)\relax
\dm*=135sp\dm*=#1\dm*\d**=#1\Lengthunit\ifdim\dm*<0pt\dm*-\dm*\fi
\multiply\dm*\r*\a*=.3\dm*\a*=#4\a*\ifdim\a*<0pt\a*-\a*\fi
\advance\dm*\a*\N*\dm*\divide\N*10000\relax
\divide\N*2\multiply\N*2\advance\N*\*one
\L*=-.25\d**\L*=#4\L*\divide\d**\N*\divide\L*\*ths
\m*\N*\divide\m*2\dm*=\the\m*5sp\l*\dm*\sm*\n*\*one\loop
\calcparab*\shl**{-\dt*}\advance\n*1\ifnum\n*<\N*\repeat}}

\def\arrarcto#1(#2,#3)[#4]{\L*=#1\Lengthunit\L*=.54\L*
\arcto#1(#2,#3)[#4]\rmov*(#2\L*,#3\L*){\d*=.457\L*\d*=#4\d*\d**-\d*
\rmov*(#3\d**,#2\d*){\arrow.02(#2,#3)}}}

\def\dasharcto#1(#2,#3)[#4]{\rlap{\toks0={#2}\toks1={#3}\relax
\calcnum*#1(#2,#3)\dm*=\the\N*5sp\a*=.3\dm*\a*=#4\a*\ifdim\a*<0pt\a*-\a*\fi
\advance\dm*\a*\N*\dm*
\divide\N*20\multiply\N*2\advance\N*1\d**=#1\Lengthunit
\L*=-.25\d**\L*=#4\L*\divide\d**\N*\divide\L*\*ths
\m*\N*\divide\m*2\dm*=\the\m*5sp\l*\dm*
\sm*\n*\*one\loop\calcparab*
\shl**{-\dt*}\advance\n*1\ifnum\n*>\N*\else\calcparab*
\sh*(#2,#3){\xL*=#3\dt* \yL*=#2\dt*
\rx* \the\cos*\xL* \tmp* \the\sin*\yL* \advance\rx*\tmp*
\ry* \the\cos*\yL* \tmp* \the\sin*\xL* \advance\ry*-\tmp*
\kern\rx*\lower\ry*\hbox{\sm*}}\fi
\advance\n*1\ifnum\n*<\N*\repeat}}

\def\*shl*#1{\c*=\the\n*\d**\advance\c*#1\a**\d*\dt*\advance\d*#1\b**
\a*=\the\toks0\c*\b*=\the\toks1\d*\advance\a*-\b*
\b*=\the\toks1\c*\d*=\the\toks0\d*\advance\b*\d*
\rx* \the\cos*\a* \tmp* \the\sin*\b* \advance\rx*-\tmp*
\ry* \the\cos*\b* \tmp* \the\sin*\a* \advance\ry*\tmp*
\raise\ry*\rlap{\kern\rx*\unhcopy\spl*}}

\def\calcnormal*#1{\b**=10000sp\a**\b**\k*\n*\advance\k*-\m*
\multiply\a**\k*\divide\a**\m*\a**=#1\a**\ifdim\a**<0pt\a**-\a**\fi
\ifdim\a**>\b**\d*=.96\a**\advance\d*.4\b**
\else\d*=.96\b**\advance\d*.4\a**\fi
\d*=.01\d*\r*\d*\divide\a**\r*\divide\b**\r*
\ifnum\k*<0\a**-\a**\fi\d*=#1\d*\ifdim\d*<0pt\b**-\b**\fi
\k*\a**\a**=\the\k*\dd*\k*\b**\b**=\the\k*\dd*}

\def\wavearcto#1(#2,#3)[#4]{\rlap{\toks0={#2}\toks1={#3}\relax
\calcnum*#1(#2,#3)\c*=\the\N*5sp\a*=.4\c*\a*=#4\a*\ifdim\a*<0pt\a*-\a*\fi
\advance\c*\a*\N*\c*\divide\N*20\multiply\N*2\advance\N*-1\multiply\N*4\relax
\d**=#1\Lengthunit\dd*=.012\d**
\divide\dd*\*ths \multiply\dd*\magnitude
\ifdim\d**<0pt\d**-\d**\fi\L*=.25\d**
\divide\d**\N*\divide\dd*\N*\L*=#4\L*\divide\L*\*ths
\m*\N*\divide\m*2\dm*=\the\m*0sp\l*\dm*
\sm*\n*\*one\loop\calcnormal*{#4}\calcparab*
\*shl*{1}\advance\n*\*one\calcparab*
\*shl*{1.3}\advance\n*\*one\calcparab*
\*shl*{1}\advance\n*2\dd*-\dd*\ifnum\n*<\N*\repeat\n*\N*\shl**{0pt}}}

\def\triangarcto#1(#2,#3)[#4]{\rlap{\toks0={#2}\toks1={#3}\relax
\calcnum*#1(#2,#3)\c*=\the\N*5sp\a*=.4\c*\a*=#4\a*\ifdim\a*<0pt\a*-\a*\fi
\advance\c*\a*\N*\c*\divide\N*20\multiply\N*2\advance\N*-1\multiply\N*2\relax
\d**=#1\Lengthunit\dd*=.012\d**
\divide\dd*\*ths \multiply\dd*\magnitude
\ifdim\d**<0pt\d**-\d**\fi\L*=.25\d**
\divide\d**\N*\divide\dd*\N*\L*=#4\L*\divide\L*\*ths
\m*\N*\divide\m*2\dm*=\the\m*0sp\l*\dm*
\sm*\n*\*one\loop\calcnormal*{#4}\calcparab*
\*shl*{1}\advance\n*2\dd*-\dd*\ifnum\n*<\N*\repeat\n*\N*\shl**{0pt}}}

\def\hr*#1{\L*=\xscale\Lengthunit\ifnum
\angle**=0\clap{\vrule width#1\L* height.1pt}\else
\L*=#1\L*\L*=.5\L*\rmov*(-\L*,0pt){\sm*}\rmov*(\L*,0pt){\sl*}\fi}

\def\shade#1[#2]{\rlap{\Lengthunit=#1\Lengthunit
\special{em:linewidth .001pt}\relax
\mov(0,#2.05){\hr*{.994}}\mov(0,#2.1){\hr*{.980}}\relax
\mov(0,#2.15){\hr*{.953}}\mov(0,#2.2){\hr*{.916}}\relax
\mov(0,#2.25){\hr*{.867}}\mov(0,#2.3){\hr*{.798}}\relax
\mov(0,#2.35){\hr*{.715}}\mov(0,#2.4){\hr*{.603}}\relax
\mov(0,#2.45){\hr*{.435}}\special{em:linewidth \the\linwid*}}}

\def\dshade#1[#2]{\rlap{\special{em:linewidth .001pt}\relax
\Lengthunit=#1\Lengthunit\if#2-\def\t*{+}\else\def\t*{-}\fi
\mov(0,\t*.025){\relax
\mov(0,#2.05){\hr*{.995}}\mov(0,#2.1){\hr*{.988}}\relax
\mov(0,#2.15){\hr*{.969}}\mov(0,#2.2){\hr*{.937}}\relax
\mov(0,#2.25){\hr*{.893}}\mov(0,#2.3){\hr*{.836}}\relax
\mov(0,#2.35){\hr*{.760}}\mov(0,#2.4){\hr*{.662}}\relax
\mov(0,#2.45){\hr*{.531}}\mov(0,#2.5){\hr*{.320}}\relax
\special{em:linewidth \the\linwid*}}}}

\def\vdot{\rlap{\kern-1.9pt\lower1.8pt\hbox{$\scriptstyle\bullet$}}}
\def\vtimes{\rlap{\kern-3pt\lower1.8pt\hbox{$\scriptstyle\times$}}}
\def\vDot{\rlap{\kern-2.3pt\lower2.7pt\hbox{$\bullet$}}}
\def\vTimes{\rlap{\kern-3.6pt\lower2.4pt\hbox{$\times$}}}

\def\arc(#1)[#2,#3]{{\k*=#2\l*=#3\m*=\l*
\advance\m*-6\ifnum\k*>\l*\relax\else
{\rotate(#2)\mov(#1,0){\sm*}}\loop
\ifnum\k*<\m*\advance\k*5{\rotate(\k*)\mov(#1,0){\sl*}}\repeat
{\rotate(#3)\mov(#1,0){\sl*}}\fi}}

\def\dasharc(#1)[#2,#3]{{\k**=#2\n*=#3\advance\n*-1\advance\n*-\k**
\L*=1000sp\L*#1\L* \multiply\L*\n* \multiply\L*\Nhalfperiods
\divide\L*57\N*\L* \divide\N*2000\ifnum\N*=0\N*1\fi
\r*\n*  \divide\r*\N* \ifnum\r*<2\r*2\fi
\m**\r* \divide\m**2 \l**\r* \advance\l**-\m** \N*\n* \divide\N*\r*
\k**\r* \multiply\k**\N* \dn*\n* \advance\dn*-\k** \divide\dn*2\advance\dn*\*one
\r*\l** \divide\r*2\advance\dn*\r* \advance\N*-2\k**#2\relax
\ifnum\l**<6{\rotate(#2)\mov(#1,0){\sm*}}\advance\k**\dn*
{\rotate(\k**)\mov(#1,0){\sl*}}\advance\k**\m**
{\rotate(\k**)\mov(#1,0){\sm*}}\loop
\advance\k**\l**{\rotate(\k**)\mov(#1,0){\sl*}}\advance\k**\m**
{\rotate(\k**)\mov(#1,0){\sm*}}\advance\N*-1\ifnum\N*>0\repeat
{\rotate(#3)\mov(#1,0){\sl*}}\else\advance\k**\dn*
\arc(#1)[#2,\k**]\loop\advance\k**\m** \r*\k**
\advance\k**\l** {\arc(#1)[\r*,\k**]}\relax
\advance\N*-1\ifnum\N*>0\repeat
\advance\k**\m**\arc(#1)[\k**,#3]\fi}}

\def\triangarc#1(#2)[#3,#4]{{\k**=#3\n*=#4\advance\n*-\k**
\L*=1000sp\L*#2\L* \multiply\L*\n* \multiply\L*\Nhalfperiods
\divide\L*57\N*\L* \divide\N*1000\ifnum\N*=0\N*1\fi
\d**=#2\Lengthunit \d*\d** \divide\d*57\multiply\d*\n*
\r*\n*  \divide\r*\N* \ifnum\r*<2\r*2\fi
\m**\r* \divide\m**2 \l**\r* \advance\l**-\m** \N*\n* \divide\N*\r*
\dt*\d* \divide\dt*\N* \dt*.5\dt* \dt*#1\dt*
\divide\dt*1000\multiply\dt*\magnitude
\k**\r* \multiply\k**\N* \dn*\n* \advance\dn*-\k** \divide\dn*2\relax
\r*\l** \divide\r*2\advance\dn*\r* \advance\N*-1\k**#3\relax
{\rotate(#3)\mov(#2,0){\sm*}}\advance\k**\dn*
{\rotate(\k**)\mov(#2,0){\sl*}}\advance\k**-\m**\advance\l**\m**\loop\dt*-\dt*
\d*\d** \advance\d*\dt*
\advance\k**\l**{\rotate(\k**)\rmov*(\d*,0pt){\sl*}}%
\advance\N*-1\ifnum\N*>0\repeat\advance\k**\m**
{\rotate(\k**)\mov(#2,0){\sl*}}{\rotate(#4)\mov(#2,0){\sl*}}}}

\def\wavearc#1(#2)[#3,#4]{{\k**=#3\n*=#4\advance\n*-\k**
\L*=4000sp\L*#2\L* \multiply\L*\n* \multiply\L*\Nhalfperiods
\divide\L*57\N*\L* \divide\N*1000\ifnum\N*=0\N*1\fi
\d**=#2\Lengthunit \d*\d** \divide\d*57\multiply\d*\n*
\r*\n*  \divide\r*\N* \ifnum\r*=0\r*1\fi
\m**\r* \divide\m**2 \l**\r* \advance\l**-\m** \N*\n* \divide\N*\r*
\dt*\d* \divide\dt*\N* \dt*.7\dt* \dt*#1\dt*
\divide\dt*1000\multiply\dt*\magnitude
\k**\r* \multiply\k**\N* \dn*\n* \advance\dn*-\k** \divide\dn*2\relax
\divide\N*4\advance\N*-1\k**#3\relax
{\rotate(#3)\mov(#2,0){\sm*}}\advance\k**\dn*
{\rotate(\k**)\mov(#2,0){\sl*}}\advance\k**-\m**\advance\l**\m**\loop\dt*-\dt*
\d*\d** \advance\d*\dt* \dd*\d** \advance\dd*1.3\dt*
\advance\k**\r*{\rotate(\k**)\rmov*(\d*,0pt){\sl*}}\relax
\advance\k**\r*{\rotate(\k**)\rmov*(\dd*,0pt){\sl*}}\relax
\advance\k**\r*{\rotate(\k**)\rmov*(\d*,0pt){\sl*}}\relax
\advance\k**\r*
\advance\N*-1\ifnum\N*>0\repeat\advance\k**\m**
{\rotate(\k**)\mov(#2,0){\sl*}}{\rotate(#4)\mov(#2,0){\sl*}}}}

\def\gmov*#1(#2,#3)#4{\rlap{\L*=#1\Lengthunit
\xL*=#2\L* \yL*=#3\L*
\rx* \gcos*\xL* \tmp* \gsin*\yL* \advance\rx*-\tmp*
\ry* \gcos*\yL* \tmp* \gsin*\xL* \advance\ry*\tmp*
\rx*=\xscale\rx* \ry*=\yscale\ry*
\xL* \the\cos*\rx* \tmp* \the\sin*\ry* \advance\xL*-\tmp*
\yL* \the\cos*\ry* \tmp* \the\sin*\rx* \advance\yL*\tmp*
\kern\xL*\raise\yL*\hbox{#4}}}

\def\rgmov*(#1,#2)#3{\rlap{\xL*#1\yL*#2\relax
\rx* \gcos*\xL* \tmp* \gsin*\yL* \advance\rx*-\tmp*
\ry* \gcos*\yL* \tmp* \gsin*\xL* \advance\ry*\tmp*
\rx*=\xscale\rx* \ry*=\yscale\ry*
\xL* \the\cos*\rx* \tmp* \the\sin*\ry* \advance\xL*-\tmp*
\yL* \the\cos*\ry* \tmp* \the\sin*\rx* \advance\yL*\tmp*
\kern\xL*\raise\yL*\hbox{#3}}}

\def\Earc(#1)[#2,#3][#4,#5]{{\k*=#2\l*=#3\m*=\l*
\advance\m*-6\ifnum\k*>\l*\relax\else\def\xscale{#4}\def\yscale{#5}\relax
{\angle**0\rotate(#2)}\gmov*(#1,0){\sm*}\loop
\ifnum\k*<\m*\advance\k*5\relax
{\angle**0\rotate(\k*)}\gmov*(#1,0){\sl*}\repeat
{\angle**0\rotate(#3)}\gmov*(#1,0){\sl*}\relax
\def\xscale{1}\def\yscale{1}\fi}}

\def\dashEarc(#1)[#2,#3][#4,#5]{{\k**=#2\n*=#3\advance\n*-1\advance\n*-\k**
\L*=1000sp\L*#1\L* \multiply\L*\n* \multiply\L*\Nhalfperiods
\divide\L*57\N*\L* \divide\N*2000\ifnum\N*=0\N*1\fi
\r*\n*  \divide\r*\N* \ifnum\r*<2\r*2\fi
\m**\r* \divide\m**2 \l**\r* \advance\l**-\m** \N*\n* \divide\N*\r*
\k**\r*\multiply\k**\N* \dn*\n* \advance\dn*-\k** \divide\dn*2\advance\dn*\*one
\r*\l** \divide\r*2\advance\dn*\r* \advance\N*-2\k**#2\relax
\ifnum\l**<6\def\xscale{#4}\def\yscale{#5}\relax
{\angle**0\rotate(#2)}\gmov*(#1,0){\sm*}\advance\k**\dn*
{\angle**0\rotate(\k**)}\gmov*(#1,0){\sl*}\advance\k**\m**
{\angle**0\rotate(\k**)}\gmov*(#1,0){\sm*}\loop
\advance\k**\l**{\angle**0\rotate(\k**)}\gmov*(#1,0){\sl*}\advance\k**\m**
{\angle**0\rotate(\k**)}\gmov*(#1,0){\sm*}\advance\N*-1\ifnum\N*>0\repeat
{\angle**0\rotate(#3)}\gmov*(#1,0){\sl*}\def\xscale{1}\def\yscale{1}\else
\advance\k**\dn* \Earc(#1)[#2,\k**][#4,#5]\loop\advance\k**\m** \r*\k**
\advance\k**\l** {\Earc(#1)[\r*,\k**][#4,#5]}\relax
\advance\N*-1\ifnum\N*>0\repeat
\advance\k**\m**\Earc(#1)[\k**,#3][#4,#5]\fi}}

\def\triangEarc#1(#2)[#3,#4][#5,#6]{{\k**=#3\n*=#4\advance\n*-\k**
\L*=1000sp\L*#2\L* \multiply\L*\n* \multiply\L*\Nhalfperiods
\divide\L*57\N*\L* \divide\N*1000\ifnum\N*=0\N*1\fi
\d**=#2\Lengthunit \d*\d** \divide\d*57\multiply\d*\n*
\r*\n*  \divide\r*\N* \ifnum\r*<2\r*2\fi
\m**\r* \divide\m**2 \l**\r* \advance\l**-\m** \N*\n* \divide\N*\r*
\dt*\d* \divide\dt*\N* \dt*.5\dt* \dt*#1\dt*
\divide\dt*1000\multiply\dt*\magnitude
\k**\r* \multiply\k**\N* \dn*\n* \advance\dn*-\k** \divide\dn*2\relax
\r*\l** \divide\r*2\advance\dn*\r* \advance\N*-1\k**#3\relax
\def\xscale{#5}\def\yscale{#6}\relax
{\angle**0\rotate(#3)}\gmov*(#2,0){\sm*}\advance\k**\dn*
{\angle**0\rotate(\k**)}\gmov*(#2,0){\sl*}\advance\k**-\m**
\advance\l**\m**\loop\dt*-\dt* \d*\d** \advance\d*\dt*
\advance\k**\l**{\angle**0\rotate(\k**)}\rgmov*(\d*,0pt){\sl*}\relax
\advance\N*-1\ifnum\N*>0\repeat\advance\k**\m**
{\angle**0\rotate(\k**)}\gmov*(#2,0){\sl*}\relax
{\angle**0\rotate(#4)}\gmov*(#2,0){\sl*}\def\xscale{1}\def\yscale{1}}}

\def\waveEarc#1(#2)[#3,#4][#5,#6]{{\k**=#3\n*=#4\advance\n*-\k**
\L*=4000sp\L*#2\L* \multiply\L*\n* \multiply\L*\Nhalfperiods
\divide\L*57\N*\L* \divide\N*1000\ifnum\N*=0\N*1\fi
\d**=#2\Lengthunit \d*\d** \divide\d*57\multiply\d*\n*
\r*\n*  \divide\r*\N* \ifnum\r*=0\r*1\fi
\m**\r* \divide\m**2 \l**\r* \advance\l**-\m** \N*\n* \divide\N*\r*
\dt*\d* \divide\dt*\N* \dt*.7\dt* \dt*#1\dt*
\divide\dt*1000\multiply\dt*\magnitude
\k**\r* \multiply\k**\N* \dn*\n* \advance\dn*-\k** \divide\dn*2\relax
\divide\N*4\advance\N*-1\k**#3\def\xscale{#5}\def\yscale{#6}\relax
{\angle**0\rotate(#3)}\gmov*(#2,0){\sm*}\advance\k**\dn*
{\angle**0\rotate(\k**)}\gmov*(#2,0){\sl*}\advance\k**-\m**
\advance\l**\m**\loop\dt*-\dt*
\d*\d** \advance\d*\dt* \dd*\d** \advance\dd*1.3\dt*
\advance\k**\r*{\angle**0\rotate(\k**)}\rgmov*(\d*,0pt){\sl*}\relax
\advance\k**\r*{\angle**0\rotate(\k**)}\rgmov*(\dd*,0pt){\sl*}\relax
\advance\k**\r*{\angle**0\rotate(\k**)}\rgmov*(\d*,0pt){\sl*}\relax
\advance\k**\r*
\advance\N*-1\ifnum\N*>0\repeat\advance\k**\m**
{\angle**0\rotate(\k**)}\gmov*(#2,0){\sl*}\relax
{\angle**0\rotate(#4)}\gmov*(#2,0){\sl*}\def\xscale{1}\def\yscale{1}}}

\newcount\CatcodeOfAtSign
\CatcodeOfAtSign=\the\catcode`\@
\catcode`\@=11
\def\@arc#1[#2][#3]{\rlap{\Lengthunit=#1\Lengthunit
\sm*\l*arc(#2.1914,#3.0381)[#2][#3]\relax
\mov(#2.1914,#3.0381){\l*arc(#2.1622,#3.1084)[#2][#3]}\relax
\mov(#2.3536,#3.1465){\l*arc(#2.1084,#3.1622)[#2][#3]}\relax
\mov(#2.4619,#3.3086){\l*arc(#2.0381,#3.1914)[#2][#3]}}}

\def\dash@arc#1[#2][#3]{\rlap{\Lengthunit=#1\Lengthunit
\d*arc(#2.1914,#3.0381)[#2][#3]\relax
\mov(#2.1914,#3.0381){\d*arc(#2.1622,#3.1084)[#2][#3]}\relax
\mov(#2.3536,#3.1465){\d*arc(#2.1084,#3.1622)[#2][#3]}\relax
\mov(#2.4619,#3.3086){\d*arc(#2.0381,#3.1914)[#2][#3]}}}

\def\wave@arc#1[#2][#3]{\rlap{\Lengthunit=#1\Lengthunit
\w*lin(#2.1914,#3.0381)\relax
\mov(#2.1914,#3.0381){\w*lin(#2.1622,#3.1084)}\relax
\mov(#2.3536,#3.1465){\w*lin(#2.1084,#3.1622)}\relax
\mov(#2.4619,#3.3086){\w*lin(#2.0381,#3.1914)}}}

\def\bezier#1(#2,#3)(#4,#5)(#6,#7){\N*#1\l*\N* \advance\l*\*one
\d* #4\Lengthunit \advance\d* -#2\Lengthunit \multiply\d* \*two
\b* #6\Lengthunit \advance\b* -#2\Lengthunit
\advance\b*-\d* \divide\b*\N*
\d** #5\Lengthunit \advance\d** -#3\Lengthunit \multiply\d** \*two
\b** #7\Lengthunit \advance\b** -#3\Lengthunit
\advance\b** -\d** \divide\b**\N*
\mov(#2,#3){\sm*{\loop\ifnum\m*<\l*
\a*\m*\b* \advance\a*\d* \divide\a*\N* \multiply\a*\m*
\a**\m*\b** \advance\a**\d** \divide\a**\N* \multiply\a**\m*
\rmov*(\a*,\a**){\unhcopy\spl*}\advance\m*\*one\repeat}}}

\catcode`\*=12

\newcount\n@ast
\def\n@ast@#1{\n@ast0\relax\get@ast@#1\end}
\def\get@ast@#1{\ifx#1\end\let\next\relax\else
\ifx#1*\advance\n@ast1\fi\let\next\get@ast@\fi\next}

\newif\if@up \newif\if@dwn
\def\up@down@#1{\@upfalse\@dwnfalse
\if#1u\@uptrue\fi\if#1U\@uptrue\fi\if#1+\@uptrue\fi
\if#1d\@dwntrue\fi\if#1D\@dwntrue\fi\if#1-\@dwntrue\fi}

\def\halfcirc#1(#2)[#3]{{\Lengthunit=#2\Lengthunit\up@down@{#3}\relax
\if@up\mov(0,.5){\@arc[-][-]\@arc[+][-]}\fi
\if@dwn\mov(0,-.5){\@arc[-][+]\@arc[+][+]}\fi
\def\lft{\mov(0,.5){\@arc[-][-]}\mov(0,-.5){\@arc[-][+]}}\relax
\def\rght{\mov(0,.5){\@arc[+][-]}\mov(0,-.5){\@arc[+][+]}}\relax
\if#3l\lft\fi\if#3L\lft\fi\if#3r\rght\fi\if#3R\rght\fi
\n@ast@{#1}\relax
\ifnum\n@ast>0\if@up\shade[+]\fi\if@dwn\shade[-]\fi\fi
\ifnum\n@ast>1\if@up\dshade[+]\fi\if@dwn\dshade[-]\fi\fi}}

\def\halfdashcirc(#1)[#2]{{\Lengthunit=#1\Lengthunit\up@down@{#2}\relax
\if@up\mov(0,.5){\dash@arc[-][-]\dash@arc[+][-]}\fi
\if@dwn\mov(0,-.5){\dash@arc[-][+]\dash@arc[+][+]}\fi
\def\lft{\mov(0,.5){\dash@arc[-][-]}\mov(0,-.5){\dash@arc[-][+]}}\relax
\def\rght{\mov(0,.5){\dash@arc[+][-]}\mov(0,-.5){\dash@arc[+][+]}}\relax
\if#2l\lft\fi\if#2L\lft\fi\if#2r\rght\fi\if#2R\rght\fi}}

\def\halfwavecirc(#1)[#2]{{\Lengthunit=#1\Lengthunit\up@down@{#2}\relax
\if@up\mov(0,.5){\wave@arc[-][-]\wave@arc[+][-]}\fi
\if@dwn\mov(0,-.5){\wave@arc[-][+]\wave@arc[+][+]}\fi
\def\lft{\mov(0,.5){\wave@arc[-][-]}\mov(0,-.5){\wave@arc[-][+]}}\relax
\def\rght{\mov(0,.5){\wave@arc[+][-]}\mov(0,-.5){\wave@arc[+][+]}}\relax
\if#2l\lft\fi\if#2L\lft\fi\if#2r\rght\fi\if#2R\rght\fi}}

\catcode`\*=11

\def\Circle#1(#2){\halfcirc#1(#2)[u]\halfcirc#1(#2)[d]\n@ast@{#1}\relax
\ifnum\n@ast>0\L*=\xscale\Lengthunit
\ifnum\angle**=0\clap{\vrule width#2\L* height.1pt}\else
\L*=#2\L*\L*=.5\L*\special{em:linewidth .001pt}\relax
\rmov*(-\L*,0pt){\sm*}\rmov*(\L*,0pt){\sl*}\relax
\special{em:linewidth \the\linwid*}\fi\fi}

\catcode`\*=12

\def\wavecirc(#1){\halfwavecirc(#1)[u]\halfwavecirc(#1)[d]}

\def\dashcirc(#1){\halfdashcirc(#1)[u]\halfdashcirc(#1)[d]}

\def\xscale{1}
\def\yscale{1}

\def\Ellipse#1(#2)[#3,#4]{\def\xscale{#3}\def\yscale{#4}\relax
\Circle#1(#2)\def\xscale{1}\def\yscale{1}}

\def\dashEllipse(#1)[#2,#3]{\def\xscale{#2}\def\yscale{#3}\relax
\dashcirc(#1)\def\xscale{1}\def\yscale{1}}

\def\waveEllipse(#1)[#2,#3]{\def\xscale{#2}\def\yscale{#3}\relax
\wavecirc(#1)\def\xscale{1}\def\yscale{1}}

\def\halfEllipse#1(#2)[#3][#4,#5]{\def\xscale{#4}\def\yscale{#5}\relax
\halfcirc#1(#2)[#3]\def\xscale{1}\def\yscale{1}}

\def\halfdashEllipse(#1)[#2][#3,#4]{\def\xscale{#3}\def\yscale{#4}\relax
\halfdashcirc(#1)[#2]\def\xscale{1}\def\yscale{1}}

\def\halfwaveEllipse(#1)[#2][#3,#4]{\def\xscale{#3}\def\yscale{#4}\relax
\halfwavecirc(#1)[#2]\def\xscale{1}\def\yscale{1}}

\catcode`\@=\the\CatcodeOfAtSign

The study of strong interaction contributions to the energy spectra
of hydrogen-like systems and lepton anomalous magnetic moments (AMM)
is considered lately as an important problem which has large
practical meaning for testing the Standard Model \cite{EGS,N2002}.
One of such contributions is determined by hadronic vacuum polarization
(HVP) \cite{C,GKF}. Presently there are at least two tasks where
the HVP effects can be tested on the experiment. The measurement of muon
anomalous magnetic moment in the new E821 experiment at the Brookhaven
National Laboratory (BNL) was done with extremely high accuracy \cite{E821,N2002}:
\begin{equation}
{\rm a_\mu=116~592~023~(151)\times 10^{-11}},
\end{equation}
which allows to perform a check of the hadronic contribution in the Standard
Model not only at one-loop level, but also with two-loop accuracy. In a Los
Alamos experiment aimed at measuring the hyperfine splitting (HFS) of the muonium
ground state, the accuracy of the measurement reached a few tens of
Hertz \cite{Liu}:
\begin{equation}
{\rm \Delta\nu^{HFS}(Mu)=4~463~302~765~(53) Hz}.
\end{equation}
This requires taking into account both higher
order contributions in $\alpha$ ($\alpha$ is the fine structure constant) and
the contribution from the HVP of different order
\cite{FKM,KS,N,CEK} to the muonium hyperfine splitting. Consistent calculation
of the HVP
contribution to the photon Green's function can be carried out on high level
of the accuracy, because hadronic spectral function ${\rho^h}$ is connected with
the cross section of ${\rm e^+e^-}$ annihilation into hadrons measured
experimentally. A completely different situation occurs in the case
of the hadronic light-by-light scattering contribution to the HFS in hydrogenic
atoms or to the muon AMM \cite{KNO,HKS,Kuraev,BEP1} (see diagrams in fig.1,2).
The contribution of such diagrams to the energy spectrum or lepton AMM is
determined by the tensor of virtual $\gamma^\ast\gamma^\ast$ scattering
containing eight structure functions. Reliable experimental data about
these structure functions are not available at present. In the low energy region
${\rm\gamma^\ast\gamma^\ast}$ interaction may be studied by means of effective
field theory of interacting photons and hadrons \cite{Knecht}. The calculation of such
hadronic contributions to muon AMM in the pion pole terms approximation
was carried out in Ref. \cite{KN}. As a result the difference between
the theoretical value for the muon AMM and experimental data was reduced to about
1 standard deviation \cite{N2002,HK2001,BCM,BEP2}. In this work we calculated the
contribution of pseudoscalar pole terms to the muonium HFS \cite{FKM}. Corresponding
diagrams,
shown in fig.1,2 describe a part of ${\rm O(\alpha)}$ corrections to the
leading hadronic contribution in the muonium HFS. The investigation of higher order
effects on $\alpha$ to the muonium ground state hyperfine splitting allows
to determine the muon-to-electron mass ratio from  experimentally
measured value (2) \cite{EGS,EGS1,Nio,Hill}:
\begin{equation}
{\rm \frac{m_\mu}{m_e}=206.768~279~8(43),~~~\delta=2\cdot 10^{-8}.}
\end{equation}
Such indirect determination of the muon electron mass ratio is almost six times
more accurate than the best experimental value for ${\rm m_\mu/m_e}$ \cite{Liu}.
Further improvement of the accuracy in (3) is connected with the calculation
of a new theoretical contributions to ${\rm\Delta\nu_{HFS}^{th}(Mu)}$
of order 10 Hz, containing small parameters ${\rm\alpha^4 E_F\approx}$ 13 Hz,
${\rm\alpha^3\frac{m_e}{m_\mu}E_F\approx}$ 8 Hz, ${\rm\alpha^2(m_e/m_\mu)^2E_F
\approx}$ 6 Hz which can be enhanced by the presence of logarithmic factors
${\rm \ln\alpha^{-1}}$ = 4.92, ${\rm\ln(m_\mu/m_e)}$ = 5.33.
\begin{figure*}[t!]
\newdimen\Lengthunit       \Lengthunit  = 1.8cm
\magnitude=1500
\GRAPH(hsize=15){
\mov(0,0){\lin(3,0)}%
\mov(3.2,0){${\rm\mu^+}$}%
\mov(3.2,2){${\rm e^-}$}%
\mov(4,0){\lin(2,0)}%
\mov(5,-0.4){a}%
\mov(7,0){\lin(2,0)}%
\mov(8,-0.4){b}%
\mov(0,2){\lin(3,0)}%
\mov(4,2){\lin(2,0)}%
\mov(7,2){\lin(2,0)}%
\mov(1.5,1.){\Circle*(1.0)}%
\mov(0.5,0){\wavelin(0.65,0.65)}%
\mov(2.5,0){\wavelin(-0.65,0.65)}%
\mov(0.5,2.){\wavelin(0.65,-0.65)}%
\mov(2.5,2){\wavelin(-0.65,-0.65)}%
\mov(3.5,1){=}%
\mov(6.5,1.){+}%
\mov(4.5,1.){\Circle*(0.2)}%
\mov(5.5,1.){\Circle*(0.2)}%
\mov(7.5,1.){\Circle*(0.2)}%
\mov(8.5,1.){\Circle*(0.2)}%
\mov(4.5,1.1){\lin(1,0)}%
\mov(4.5,0.9){\lin(1,0)}%
\mov(7.5,1.1){\lin(1,0)}%
\mov(7.5,0.9){\lin(1,0)}%
\mov(4.5,1.1){\wavelin(-0.25,0.9)}%
\mov(4.5,0.9){\wavelin(-0.25,-0.9)}%
\mov(5.5,1.1){\wavelin(0.25,0.9)}%
\mov(5.5,0.9){\wavelin(0.25,-0.9)}%
\mov(7.5,0.9){\wavelin(-0.25,-0.9)}%
\mov(8.5,0.9){\wavelin(0.25,-0.9)}%
\mov(8.5,1.1){\wavelin(-1.,0.9)}%
\mov(7.5,1.1){\wavelin(1.,0.9)}%
}
\caption{Pseudoscalar pole terms giving rise the contribution of order
${\rm\alpha^3\left(\frac{m_em_\mu}{F_\pi^2}\right)E_F}$ to the muonium HFS.}
\end{figure*}

\begin{figure*}[t!]
\newdimen\Lengthunit       \Lengthunit  = 1.7cm
\magnitude=1500
\GRAPH(hsize=15){
\mov(0,0){\lin(2,0)}%
\mov(2.2,0){${\rm\mu^+}$}%
\mov(2.2,2){${\rm e^-}$}%
\mov(2.5,0){\lin(2,0)}%
\mov(3.5,-0.4){a}%
\mov(5,0){\lin(2,0)}%
\mov(6,-0.4){b}%
\mov(7.5,0){\lin(2,0)}%
\mov(8.5,-0.4){c}%
\mov(0,2){\lin(2,0)}%
\mov(2.5,2){\lin(2,0)}%
\mov(5,2){\lin(2,0)}%
\mov(7.5,2){\lin(2,0)}%
\mov(1,1){\Ellipse*(1.4)[1.,0.7]}%
\mov(1,2){\wavelin(0.,-0.5)}%
\mov(1,0){\wavelin(0,0.5)}%
\mov(0.5,0){\wavelin(0,0.65)}%
\mov(1.5,0){\wavelin(0,0.65)}%
\mov(2.25,1){=}%
\mov(4.75,1){+}%
\mov(7.25,1){+}%
\mov(3.,1.){\Circle*(0.1)}%
\mov(3.,0.95){\lin(1.,0)}%
\mov(3.,1.05){\lin(1.,0.)}%
\mov(3.,0.95){\wavelin(0.3,-0.95)}%
\mov(3.,0.95){\wavelin(-0.3,-0.95)}%
\mov(4.,1.){\Circle*(0.1)}%
\mov(4.,0){\wavelin(0,0.95)}%
\mov(4.,2.){\wavelin(0.,-0.95)}%
\mov(5.5,1.){\Circle*(0.1)}%
\mov(6.5,1.){\Circle*(0.1)}%
\mov(6.5,0.95){\lin(-1.,0)}%
\mov(6.5,1.05){\lin(-1.,0.)}%
\mov(6.5,0.95){\wavelin(0.3,-0.95)}%
\mov(6.5,0.95){\wavelin(-0.3,-0.95)}%
\mov(5.5,0){\wavelin(0,0.95)}%
\mov(5.5,2.){\wavelin(0.,-0.95)}%
\mov(8.,1.){\Circle*(0.1)}%
\mov(9.,1.){\Circle*(0.1)}%
\mov(8.,0.95){\lin(1.,0)}%
\mov(8.,1.05){\lin(1.,0.)}%
\mov(8.,0.95){\wavelin(1.3,-0.95)}%
\mov(8.,0.95){\wavelin(-0.3,-0.95)}%
\mov(8.5,0){\wavelin(0.5,0.95)}%
\mov(9.,2.){\wavelin(0.,-0.95)}%
}
\caption{Pseudoscalar pole terms giving rise the contribution of order
${\rm\alpha^3\left(\frac{m_\mu M_V}{F_\pi^2}\right)E_F}$ to
the muonium HFS.}
\end{figure*}

The effective vertex for the interaction of a ${\rm \pi^0}$ meson
( or other pseudoscalar mesons $\eta$, $\eta'$) and virtual
photons can be expressed by means of the transition form factor
${\rm F_{\pi^0\gamma^\ast\gamma^\ast}(k_1^2,k_2^2)}$:
\begin{equation}
{\rm V^{\mu\nu}(k_1,k_2)=i\epsilon^{\mu\nu\alpha\beta}k_{1~\alpha}k_{2~\beta}
\frac{\alpha}{\pi F_\pi}F_{\pi^0\gamma^\ast\gamma^\ast}(k_1^2,k_2^2)},
\end{equation}
where ${\rm F_\pi=0.0924}$ Gev is the pion decay constant, ${\rm k_1, k_2}$
are the photon four momenta. Different model representations for the
transition form factor ${\rm F_{\pi^0\gamma^\ast\gamma^\ast}(k_1^2,k_2^2)}$
exist. All of them can be used for the numerical estimation of the corrections
to the muonium HFS. Firstly consider corrections which are determined by the
diagrams in fig.1. In this case the photons interacting with one particle
(electron or muon) possess the momenta ${\rm k_{1,2}}$ and ${\rm (-k_{1,2})}$.
The Feynman amplitudes containing the photon momenta ${\rm k_{1,2}}$ and
${\rm (-k_{1,2}+q_{1,2}-p_{1,2})}$ (${\rm p_{1,2}}$ and ${\rm q_{1,2}}$
are four momenta of particles ${\rm e^-}$, ${\rm \mu^+}$ in the initial and
final states respectively) in one vertex (4) can be omitted since they determine the
contribution to HFS of higher order in $\alpha$.

To construct the HFS part of the quasipotential in the
${\rm e^-\mu^+}$ system we used the operators projecting
onto the states with the spin S=0 and  S=1 respectively:
\begin{equation}
{\rm \hat \Pi_{S=0}=\left[u(p_1)\bar v(-p_2)\right]_{S=0}=\frac{1+\gamma^0}{2\sqrt{2}}
\gamma_5,~~~\hat \Pi_{S=1}=\left[u(p_1)\bar v(-p_2)\right]_{S=1} =
\frac{1+\gamma^0}{2\sqrt{2}}\hat\epsilon,}
\end{equation}
where ${\rm\epsilon^\mu}$ is the muonium polarization vector for the state
${\rm ^3S_1}$, ${\rm\hat\epsilon=\epsilon^\mu\gamma_\mu}$.
To obtain the contribution of the Feynman amplitude M to the energy
spectrum we must calculate the common trace:
${\rm Tr\left[\hat\Pi^+M\hat \Pi\right]}$. The diagrams with direct photons
in fig.1 (a) give the following contribution to the muonium HFS:
\begin{equation}
{\rm \Delta E_1^{HFS}=-E_F\frac{16\alpha^3m_em_\mu}{F_\pi^2\pi n^3}\int\frac{d^4k_1}
{i(2\pi)^4}\int\frac{d^4k_2}{i(2\pi)^4}\frac{f_1(k_1,k_2)}{(k_1^2)^2(k_2^2)^2}
\frac{\left[F_{\pi^0\gamma^\ast\gamma^\ast}(k_1^2,k_2^2)\right]^2}
{\left(k_1^2-2k_1^0\frac{m_e}{M_V}\right)\left(k_2^2-2k_2^0\frac{m_\mu}
{M_V}\right)\left[(k_1+k_2)^2-\frac{m_\pi^2}{M_V^2}\right],}}
\end{equation}
where the Fermi energy ${\rm E_F=8\mu^3(Z\alpha)^4/m_em_\mu}$, ${\rm\mu=m_em_\mu/
(m_e+m_\mu)}$ is reduced mass, n is the principal quantum number,
\begin{equation}
{\rm f_1(k_1,k_2)=-2k_1^{0~2}k_2^2(k_1k_2)+k_1^0k_2^0[k_1^2k_2^2+(k_1k_2)^2]+
(k_1k_2)[k_1^2k_2^2-(k_1k_2)^2]}.
\end{equation}
We make dimensionless the integration variables ${\rm k_1, k_2}$ using the
mass of the vector meson ${\rm M_V}$. Our function ${\rm f_1(k_1,k_2)}$ was
derived after the calculation of the trace, summing over the Lorentz indices and
averaging over the directions of the three momentum ${\rm\vec k}$ for the
states with total spin S=0 and S=1 respectively:
\begin{equation}
{\rm \epsilon^{\mu\nu\alpha\beta}k_{1~\alpha}k_{2~\beta}\epsilon^{\lambda
\sigma\rho\omega}k_{1~\rho}k_{2~\omega}Tr\left[\gamma^\lambda(\hat p_1-\hat k_1+m_e)
\gamma^\mu\hat\Pi\gamma^\nu(-\hat p_2+\hat k_2+m_\mu)\gamma^\sigma\hat\Pi^+\right]}.
\end{equation}
We employ the system FORM for this aim \cite{V}.
The contribution of the Feynman diagram in fig.1 (b) to HFS with crossed photons
has the form:
\begin{equation}
{\rm \Delta E_2^{HFS}=-E_F\frac{16\alpha^3m_em_\mu}{F_\pi^2\pi n^3}\int\frac{d^4k_1}
{i(2\pi)^4}\int\frac{d^4k_2}{i(2\pi)^4}\frac{f_1(k_1,k_2)}{(k_1^2)^2(k_2^2)^2}
\frac{\left[F_{\pi\gamma^\ast\gamma^\ast}(k_1^2,k_2^2)\right]^2}
{\left(k_1^2+2k_1^0\frac{m_e}{M_V}\right)\left(k_2^2-2k_2^0\frac{m_\mu}
{M_V}\right)\left[(k_1+k_2)^2-\frac{m_\pi^2}{M_V^2}\right],}}
\end{equation}
Corrections of lower order in the ratio of particle masses to the muonium HFS are
determined by the Feynman diagrams in fig.2. They contain the factor
of the same power in $\alpha$, because the photon propagator ${\rm 1/\vec k^2}$
enters the amplitudes together with two factors ${\rm \vec k}$ in the
numerator. One degree of ${\rm \vec k}$ is related to the anomalous part
of the electron transition current ${\rm\bar u(q_1)\frac{-1}{2m_e}
\sigma^{\mu\nu}k_\nu u(p_1)}$ and the other enters one of the vertices
of the ${\rm\pi^0\gamma^\ast\gamma^\ast}$ interaction. This does necessarily mean
that the muon transition current expressed as
\begin{equation}
{\rm j^\mu_{muon}=\bar u(q_2)\Gamma^\mu u(p_2)=\bar u(q_2)\left[\frac{q_2^\mu+
p_2^\mu}{2m_\mu}F_e-\frac{1}{2m_\mu}\sigma^{\mu\nu}k_\nu F_m\right]u(p_2)},
\end{equation}
has the contributions from diagrams in fig.2 only to the magnetic form factor
${\rm F_m}$, while ${\rm F_e}$ = 0. The sum of contributions coming from
the diagrams in fig.2 (a) and (b) can be written in the unified form if
we use the symmetry properties of integral function under changing the
variables ${\rm k_1\leftrightarrow -k_2}$:
\begin{equation}
{\rm \Delta E_3^{HFS}=E_F\frac{4\alpha^3m_\mu M_V}{F_\pi^2\pi n^3}\int\frac{d^4k_1}
{i(2\pi)^4}\int\frac{d^4k_2}{i(2\pi)^4}\frac{f_3(k_1,k_2)}{(k_1^2)(k_2^2)
(k_1+k_2)^2}
\frac{F_{\pi\gamma^\ast\gamma^\ast}(k_1^2,0)
F_{\pi\gamma^\ast\gamma^\ast}\left[k_2^2,(k_1+k_2)^2\right]}
{\left(k_1^2+2k_1^0\frac{m_e}{M_V}\right)\left(k_2^2-2k_2^0\frac{m_\mu}
{M_V}\right)\left(k_1^2-\frac{m_\pi^2}{M_V^2}\right)},}
\end{equation}
where
\begin{equation}
{\rm f_3(k_1,k_2)=-\frac{16}{3}k_1^{0~2}\frac{m_\mu}{M_V}k_2^2+\frac{16}{3}
k_1^0k_2^0\frac{m_\mu}{M_V}(k_1k_2)+k_1^0\left[\frac{16}{3}(k_1k_2)^2-
8k_1^2k_2^2\right]
+\frac{8}{3}
k_2^0k_1^2(k_1k_2)+\frac{16}{3}\frac{m_\mu}{M_V}\left[k_1^2k_2^2-(k_1k_2)^2\right]}.
\end{equation}
The contribution of the diagram in fig.2 (c) which has different structure
of the pion propagator and the transition form factors
${\rm F_{\pi^0\gamma^\ast\gamma^\ast}(k_1^2,k_2^2)}$ can be presented as follows:
\begin{equation}
{\rm \Delta E_4^{HFS}=E_F\frac{4\alpha^3m_\mu M_V}{F_\pi^2\pi n^3}\int\frac{d^4k_1}
{i(2\pi)^4}\int\frac{d^4k_2}{i(2\pi)^4}\frac{f_4(k_1,k_2)}{(k_1^2)(k_2^2)
(k_1+k_2)^2}
\frac{F_{\pi^0\gamma^\ast\gamma^\ast}\left[(k_1+k_2)^2,0\right]
F_{\pi^0\gamma^\ast\gamma^\ast}(k_1^2,k_2^2)}
{\left(k_1^2+2k_1^0\frac{m_e}{M_V}\right)\left(k_2^2-2k_2^0\frac{m_\mu}
{M_V}\right)\left[(k_1+k_2)^2-\frac{m_\pi^2}{M_V^2}\right]},}
\end{equation}
where
\begin{equation}
{\rm f_4(k_1,k_2)=-\frac{16}{3}k_1^{0~2}\frac{m_\mu}{M_V}k_2^2+\frac{16}{3}
k_1^0k_2^0\frac{m_\mu}{M_V}(k_1k_2)+
\frac{8}{3}k_1^0\left[k_1^2k_2^2+k_2^2(k_1k_2)\right]+
\frac{8}{3}\frac{m_\mu}{M_V}\left[k_1^2k_2^2-(k_1k_2)^2\right]}.
\end{equation}
Taking into account that the energy corrections (11) and (13) have the
structure ${\rm E_Fa_\mu}$ we conclude that hadronic contributions
to muon AMM from expressions (11) and (13) coincide with the relating results
of Ref. \cite{KN}. To extract the anomalous part of muon electromagnetic
current containing the form factor ${\rm F_2}$, the authors of Ref. \cite{KN}
used the projection operator of special type. The form factor
${\rm F_{\pi\gamma^\ast\gamma^\ast}(k_1^2,k_2^2)}$ describing the conversion
of two virtual photons into ${\rm\pi^0}$ meson should be calculated on the basis
of strong interaction theory. The asymptotic behaviour of this form factor
for large photon virtualities ${\rm k_{1,2}^2}$ was studied using the methods
of perturbative and nonperturbative quantum chromodynamics \cite{LB,G}. But the calculation of the
corrections to the muonium HFS requires the proper description of the form factor
${\rm F_{\pi^0\gamma^\ast\gamma^\ast}(k_1^2,k_2^2)}$ at small and intermediate
photon virtualities ${\rm k_1^2, k_2^2}$ around 1 ${\rm Gev^2}$, which are
dominant in the integrals entering ${\rm\Delta E_i^{HFS}}$. To evaluate the
obtained contributions to the muonium HFS it is convenient to use
the representation of the form factor ${\rm F_{\pi^0\gamma^\ast\gamma^\ast}(k_1^2,k_2^2)}$
based on the vector dominance model (VDM):
\begin{equation}
{\rm F_{\pi\gamma^\ast\gamma^\ast}(k_1^2,k_2^2)=
\frac{1}{\left(1-\frac{k_1^2}{\Lambda_V^2}\right)}\cdot\frac{1}{\left(1-\frac
{k_2^2}{\Lambda_V^2}\right)}},
\end{equation}
where in the case of the ${\rm\pi^0}$ contribution ${\rm \Lambda_V}$ =
${\rm M_V}$ =
${\rm M_\rho}$=0.7693 Gev is the mass of $\rho$ meson.
This double VDM ansatz which was used as a theoretical input in the CLEO data
analysis \cite{CLEO}, behaves as ${\rm 1/(k_1^2k_2^2)}$ for asympotically
large ${\rm k_1^2}$, ${\rm k_2^2}$, contradicting predictions of pQCD.
Nevertheless for reasonably small momenta ${\rm k_1^2}$, ${\rm k_2^2}$ this
model agrees with the experimental data \cite{CLEO}, so we used it to obtain
the numerical evaluation of relevant hadronic contributions.
We performed numerical integration in the expressions ${\rm\Delta E_i^{HFS}}$
using (15). As a result we obtained the summary contribution to the muonium HFS
${\rm (\Delta E_1^{HFS}}$ + ${\rm\Delta E_2^{HFS})(\pi^0)}$ coming from
the diagrams in fig.1 which is equal 0.001 Hz. Respective result for the
sum of the diagrams in
fig.2 ${\rm (\Delta E_3^{HFS}}$ + ${\rm\Delta E_4^{HFS})(\pi^0)}$ =
2.515 Hz. The employment of the VDM for the form factor ${\rm F_{\pi^0\gamma^\ast
\gamma^\ast}(k_1^2,k_2^2)}$ makes possible analytical calculation of the
integrals (11), (13), which was done in Ref. \cite{BCM} by means of the
expansion of integral functions in two parameters ${\rm\delta}$ =
${\rm (m_\pi^2-m_\mu^2)/m_\mu^2}$ and ${\rm m_\mu^2/M_V^2}$.
The particle interaction amplitudes shown in fig.1 have additional
small factor ${\rm m_e/M_\rho}$ in the comparison with the amplitudes in fig.2.
So, corresponding numerical values are very different. We note also that
the numerical value of the contribution from the diagrams in fig.2 to the muonium
HFS ${\rm a_eE_F}$
which is determined by the electron anomalous magnetic moment is extremely
small: ${\rm 10^{-4}}$ Hz.

\begin{figure}[htbp] %ORIGINAL SIZE: width=1.4TRUEIN; height=1.5TRUEIN
\vspace*{0.0cm}
\epsfxsize=0.9\textwidth
\centerline{\psfig{file=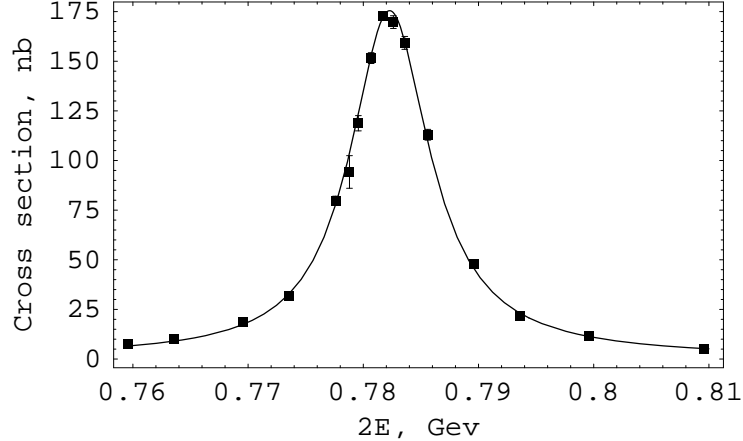,height=6.0cm,width=10.0cm}} %100 percent
\vspace*{13pt}
%\fcaption{Labeled tree {\it T}.}
\caption{Fit of the SND cross section ${\rm \sigma(e^+e^-\rightarrow\pi^0\gamma)}$}.
\end{figure}

The contributions of other pseudoscalar mesons ${\rm\eta}$ and ${\rm\eta'}$
were estimated in a similar way. To describe couplings of $\eta$ and $\eta'$
with virtual photons we used the form factor representation (4) where the
pseudoscalar decay
constants ${\rm F_\eta}$ and ${\rm F_{\eta'}}$ are related to the two photon
partial width of the resonance \cite{CLEO}:
\begin{equation}
{\rm F_{\eta,\eta'}^2=\frac{\alpha^2}{64\pi^3}\frac{M^3_{\eta,\eta'}}{\Gamma
(\eta,\eta'\rightarrow\gamma\gamma)}}.
\end{equation}
The CLEO data lead to the following values of ${\rm F_P}$ \cite{CLEO}:
${\rm F_\eta=0.0975}$ Gev, ${\rm F_{\eta'}=0.0744}$ Gev. Their fit for
transition form factors with a function defined by eq.(15)
gives the following results for the parameter ${\rm\Lambda_V}$:
${\rm\Lambda_\eta=0.774}$ Gev, ${\rm\Lambda_{\eta'}=0.859}$ Gev. The results
of numerical integrations for these two pseudoscalar mesons and total contributions
to the muonium HFS are presented in the table. Note that from the theoretical
viewpoint ${\rm F_{\eta}}$, ${\rm F_{\eta'}}$ entering in Eq. (16) should be
considered as effective decay constants due to the ${\rm\eta-\eta'}$
mixing \cite{FKS}. We considered here the pseudoscalar pole terms contributions
of order ${\rm \alpha^3E_F}$ to the muonium HFS presented in Fig. 1, 2.
Numerical values of corresponding contributions are very different.
The contributions of diagrams in Fig. 2 can be taken through the muon
anomalous magnetic moment since the value of the muonium HFS is
proportional to the product of the electron and muon total magnetic
moments. So the new contributions to the muonium HFS come from the diagrams
in Fig. 1 and are given in the first column of the table.

\begin{figure}[htbp] %ORIGINAL SIZE: width=1.4TRUEIN; height=1.5TRUEIN
\vspace*{0.0cm}
\epsfxsize=0.9\textwidth
\centerline{\psfig{file=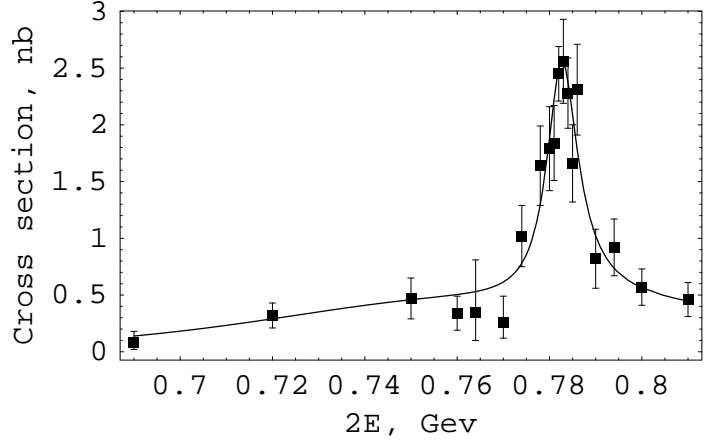,height=6.0cm,width=10.0cm}} %100 percent
\vspace*{13pt}
%\fcaption{Labeled tree {\it T}.}
\caption{Fit of the CMD-2 cross section ${\rm \sigma(e^+e^-\rightarrow\eta\gamma)}$}.
\end{figure}

\begin{figure}[htbp] %ORIGINAL SIZE: width=1.4TRUEIN; height=1.5TRUEIN
\vspace*{0.0cm}
\epsfxsize=0.9\textwidth
\centerline{\psfig{file=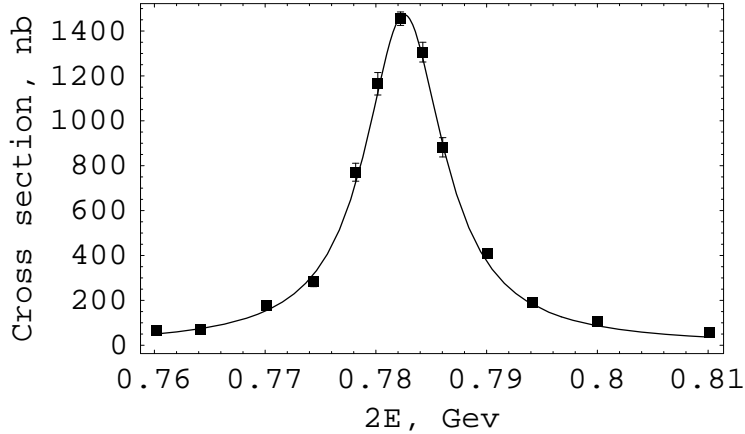,height=6.0cm,width=10.0cm}} %100 percent
\vspace*{13pt}
%\fcaption{Labeled tree {\it T}.}
\caption{Fit of the CMD-2 cross section ${\rm \sigma(e^+e^-\rightarrow
\pi^+\pi^-\pi^0)}$}.
\end{figure}

\begin{table}
\caption{Contributions of pseudoscalar mesons to light-by-light hadronic
interaction in the muonium HFS.}
\vspace{1mm}
\begin{tabular}{|c|c|c|}
P & ${\rm (\Delta E_1^{HFS}}$ + ${\rm\Delta E_2^{HFS})}$, Hz & ${\rm (\Delta E_3^{HFS}}$ +
${\rm\Delta E_4^{HFS})}$, Hz   \\  \hline
${\rm \pi^0}$& ${\rm 1.1\times 10^{-3}}$  & ${\rm 2.515}$   \\  \hline
$\eta$& ${\rm 0.2\times 10^{-3}}$ & ${\rm 0.586}$   \\  \hline
$\eta'$& ${\rm 0.1\times 10^{-3}}$  & ${\rm 0.552}$   \\  \hline
Total sum & ${\rm 1.4\times 10^{-3}}$ & 3.653  \\
\end{tabular}
\end{table}

The contributions of the annihilation processes
${\rm e^+e^-\rightarrow\rho, \omega\rightarrow\pi^0\gamma, \eta\gamma}$
to the HVP corrections to the muonium HFS have the same order of the magnitude at energies
${\rm\sqrt{s}\leq 1}$ Gev. Recently new measurements of the cross sections
of these reactions were carried out at SND and CMD-2 (Novosibirsk).
The obtained experimental data for the cross sections ${\rm\sigma(e^+e^-
\rightarrow\pi^0\gamma)}$ and ${\rm\sigma(e^+e^-\rightarrow \eta\gamma)}$
and their theoretical approximations are shown in fig.3,4.
Reactions of this type were taken into account effectively in our previous
calculations \cite{FKM} of the HVP corrections to the muonium HFS by using
the Breit-Wigner parameterization for the resonance $\omega$.
There are three important decay channels of the resonance $\omega$, which
contribute to the total decay rate with the per cent accuracy:
${\rm\omega\to\pi^+\pi^-\pi^0}$ (${\rm 88.8\%}$), ${\rm\omega\to\pi^0\gamma}$
(${\rm 8.5\%}$), ${\rm\omega\pi^+\pi^-}$ (${\rm 2.21\%}$) \cite{PDG}.
The appearance of new
experimental data for the ${\rm\sigma(e^+e^-\rightarrow
\omega\rightarrow\pi^+\pi^-\pi^0)}$, ${\rm\sigma(e^+e^-\rightarrow\rho,\omega
\rightarrow \pi^0(\eta)\gamma)}$ \cite{A3,A1,A2} allows to separate hadronic
contributions of the processes to the muonium HFS and to perform the calculation
with higher accuracy. Theoretical cross sections for these processes in the
region of $\rho, \omega$ mesons were presented in the form \cite{A3,A1,A2}:
\begin{equation}
{\rm \sigma_{\pi^0\gamma}(s)=\frac{F_{\pi^0\gamma}(s)}{s^{3/2}}|A_{\rho^0\pi^0\gamma}(s)+
A_{\omega\pi^0\gamma}(s)+A_{\phi\pi^0\gamma}(s)|^2},
\end{equation}
\begin{equation}
{\rm \sigma_{\eta\gamma}(s)=\frac{F_{\eta\gamma}(s)}{s^{3/2}}
|\sqrt{4\pi\alpha^2}C_\eta+\sum_{V=\rho,\omega,\phi,\rho'}A_{V\eta\gamma}|^2},
\end{equation}
\begin{equation}
{\rm \sigma_{3\pi}(s)=\frac{F_{3\pi}(s)}{s^{3/2}}\cdot|A_\omega+e^{i\alpha}A_\phi+
A_{bg}|^2,}
\end{equation}
\begin{equation}
{\rm A_{V\pi^0(\eta)\gamma}(s)=\frac{m_V\Gamma_V e^{i\phi_V}}{m_V^2-s-i\sqrt{s}
\Gamma_V(s)}\sqrt{\sigma_{V\pi^0(\eta)\gamma}\frac{m_V^3}{F(m_V^2)}}},
\end{equation}
where ${\rm F_{\pi^0\gamma, \eta\gamma}}$ are a phase space factors of the final
states, ${\rm \sigma_{V\pi^0(\eta)\gamma}}$ is the cross section of the
process ${\rm e^+e^-\rightarrow V
\rightarrow\pi^0(\eta)\gamma}$ at ${\rm\sqrt{s}=m_V}$, ${\rm m_V}$ is the
resonance mass, ${\rm\Gamma_V(s)}$ is its decay width at the squared c.m.
energy s, ${\rm\Gamma_V=\Gamma_V(m_V^2)}$. Three pion decay amplitudes
${\rm A_\omega}$, ${\rm A_\phi}$ have the parameterization similar to
(20) \cite{A3}. ${\rm F_{3\pi}(s)}$ is a smooth function which describes
the dynamics of ${\rm V\to\rho\pi\to\pi^+\pi^-\pi^0}$ decay \cite{KS}.
The form of the functions entering
the expressions (17)-(20) and the values of numerous parameters of the cross
sections (17) - (19) were taken also from Ref. \cite{A3,A1,A2}.
The theoretical fit (19) of the experimental data is shown in fig. 5.
To obtain the contributions to the muonium HFS we performed
numerical integration of the cross sections (17) - (19) as in the previous
work \cite{FKM} in the energy region ${\rm 0.6\leq\sqrt{s}\leq 0.81}$.
As a result we have found the following values of the corrections:
${\rm \Delta E^{HFS}_{HVP}(\rho, \omega\rightarrow\pi^0\gamma)}$ = ${\rm 1.33\pm 0.03}$
Hz, ${\rm \Delta E^{HFS}_{HVP}(\rho, \omega\rightarrow\eta\gamma)=0.042\pm
0.003}$ Hz, ${\rm\Delta E_{HVP}^{HFS}(\omega\to 3\pi)}$ = ${\rm 10.82\pm 0.33}$
Hz. So we obtain that our previous result for the $\omega$ contribution
from \cite{FKM} 12.45 Hz must be changed on 12.19 Hz.

In conclusion we can point out that among other hadronic contributions of order
${\rm \alpha^3E_F}$ there are one-loop corrections which contain in parallel
with hadronic loop also the electronic or muonic vacuum polarization
(EVP and MVP respectively). The
expressions for these contributions can be obtained using the skeleton integrals
for two photon exchange diagrams \cite{EKS,EGS1} and standard modification of the
photon propagator due to vacuum polarization and hadronic vacuum polarization.
Without going into further details we present for these corrections the
numerical values: ${\rm \Delta E^{HFS}_{EVP,HVP}}$ = 3.7 Hz,
${\rm \Delta E^{HFS}_{MVP,HVP}}$ = 1.1 Hz.

We have calculated all hadronic light-by-light corrections of order ${\rm
\alpha^3E_F}$ in the pseudoscalar pole terms approximation
and main one-loop contributions containing electron (muon) and hadronic
vacuum polarization of the same order in $\alpha$
to the muonium HFS. Combining all results presented here we find that the
theoretical value \cite{FKM} of the hadronic contribution
to the muonium ground state HFS should be increased by 8.3 Hz.

\begin{acknowledgements}
We are grateful to A.Czarnecki, S.I. Eidelman, M.I. Eides, Th. Feldmann,
S.G. Karshenboim for useful communications and discussions.
The work was performed under the financial support of the Program
"Universities of Russia" (grant UR.01.02.016) and the Ministry of Education
(grant E00-3.3-45).
\end{acknowledgements}

\end{document}